# Hydrogen electron cyclotron resonance ion sources plasma characterization based on simple optical emission spectroscopy

J. Feuchtwanger, V. Etxebarria, J. Portilla, J. Jugo, I. Badillo and I. Arredondo

*Dept. Electricidad y Electronica. Fac. Ciencia y Tecnología, Universidad del País Vasco - UPV/EHU, 48940 Leioa, Spain*

**Abstract**

Hydrogen electron cyclotron resonance ion sources plasma measurements based on simple optical emission spectroscopy on a new compact low current ion source designed and built by the authors is presented. By observing the plasma luminescence both directly and through a low-cost transmission diffraction grating, basic characterization of the Hydrogen plasma obtainable in the ion source is carried out. Through simple processing of CCD captures of these images, optimal values for the ion source relevant operation parameters, including RF power and frequency, and Hydrogen mass flow are easily obtained. Despite the simplicity of the method and its limited accuracy as compared to the use of a full standard optical spectrometric set-up, it is shown that the presented approach can cope with basic plasma diagnostic tasks as far as the successful operation of the ion source is concerned.

*Keywords:* , Ion Sources, Plasma, ECR, Optical Spectroscopy
*PACS:* 07.77.Ka, 74.25.nd, 52.50.Qt

## 1. Introduction

Electron cyclotron resonance ion sources (ECRIS) have become one of the most common means for ion production in basic research and industry for a wide range of applications because of their reliability and capability to produce multiply charged ion beams from most stable elements [1]. The performance of the ion source is fundamentally dependent on the ECR plasma production, which, in turn, depends on the microwave frequency and power used, the magnetic field's distribution and intensity, and the gas flow driven into the plasma chamber. The successful operation of an ion source requires that these variables be tuned and controlled.

The two key parameters of the plasma for ion production, density and temperature, are typically measured by means of different probes. Amongst them, Langmuir probes and its variants, such as double, capacitive, emissive or oscillation probes, are the most widely used [2]. However, these direct measurements are not always convenient or even possible during the real-time operation of the source, they are intrusive by nature and modify the plasma because they have to be inserted into the chamber, and the interpretation of their current-voltage curves can often be difficult.

On the other hand, optical spectroscopy is a very powerful method to unveil plasma physics [3]. Among its advantages it is worth mentioning that he method itself is non invasive, therefore not affecting the plasma being measured. Also, in ion sources, where RF and magnetic fields as well as high potentials are typically used, the recording of spectra is not disturbed by these. Moreover, only a line-of-sight through the plasma is required to take appropriate measurements.

For ion sources, the above mentioned key relevant parameters for ion production, such as plasma density, stability and temperature, can be deduced from optical measurements [4], which can effectively complement the existing theoretical plasma models and associated simulation codes [5] [6] which are not accurate unless based on relevant experimental data.

In this paper we present a simple optical emission spectroscopy technique developed to characterize the Hydrogen plasma obtainable in a new compact ECR ion source, named PIT30, designed and built by the authors at the University of the Basque Country (UPV/EHU). Instead of reproducing full standard optical emission methods, the approach adopted has been to simplify and avoid as much as possible the use of expensive spectrometric equipment. It is shown that if only limited accuracy is needed, optimal values for the ion source relevant operation parameters, including RF frequency, power and Hydrogen mass flow, can be easily gotten by processing simple CCD captures of the plasma luminescence both directly and through a low-cost 100 lines/mm transmission diffraction grating. The ECRIS plasma production is characterized in this way over the whole range of nominal RF power and available Hydrogen flow in the PIT30 ion source and the experimentally obtained results are discussed in detail.

Table 1: Main design parameters of PIT30 ion source

| ECRIS Parameters | |
|---|---|
| Microwave frequency | 3 GHz |
| Microwave power | <100 W |
| Gas mass flow | <5 sccm ($H_2$) |
| Magnetic field | 110 mT |
| Extraction voltage | <30 kV |
| Beam current | <100 $\mu$A |
| Beam emmitance | <0.2 $\pi$ mm mrad |

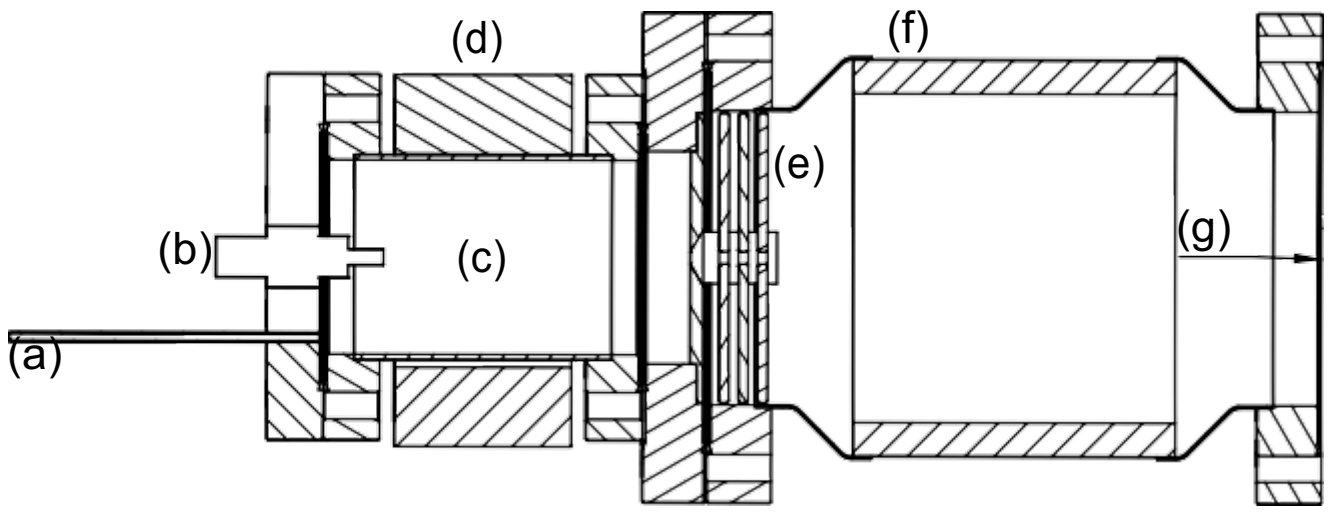

Figure 1: *Schematic diagram of the compact PIT30 ECRIS. The following components are observed: The gas inlet port (a), The coaxial RF coupler to the chamber (b), The plasma chamber itself (c), embraced by the Halbach type permanent magnet array structure (d), The extraction and focusing triplet HV electrodes (e), The ceramic insulating beam pipe (f) connecting to the diagnostics ports (g) at ground potential.*

## 2. Experimental setup

The plasma reactor where the Hydrogen plasma measurements are carried out is part of a new compact ECR ion source for low current applications developed by the authors whose main design parameters are summarized in Table 1. The ion source can be operated with gases other than Hydrogen, for instance Helium, Nitrogen and other gases. However, in this paper only Hydrogen plasma experiments will be considered.

A schematic drawing of the PIT30 ion source is shown in Fig. 1. The RF power is injected to the plasma chamber through the N coaxial port shown to the left of the Figure. This port is connected to a variable frequency RF source with a nominal central frequency of 3 GHz, connected to a compact solid state RF amplifier whose nominal power output is 100 W [1] in continuous wave. The adaptation of this RF power coupling circuit to the variable unknown plasma impedance is carried out by means of a three stub tuner inserted between the RF amplifier and the plasma chamber.

The Hydrogen gas mass flow is regulated in the range of 0 to 5 standard cubic centimeters per minute (sccm) using an Omega mass flow controler, and it is injected into the chamber though the gas inlet port. The electron cyclotron condition is met for a magnetic field given by $B = \frac{2\pi f}{e/m}$, where $f$ is the RF frequency and $e/m$ is the electron charge to mass ratio. This equation is derived fron the Larmor precession frequency. This magnetic field is provided inside the chamber through an appropriately designed Halbach type permanent magnetic structure designed to provide the ECR field in the center of the cavity for the RF frequencies used. Finally, the ions (protons) are extracted by means of an HV electrode and a triplet Einzel lens for initial electrostatic beam focusing.

The plasma measurements are carried out through the line-of-sight available from the diagnostic port shown at the far right part of Figure 1. In the simplest case, only a CCD reflex camera is used to capture the Hydrogen plasma luminescence, either directly or through a simple 100 lines/mm diffraction grating mounted as a standard filter on the camera lens. By simple processing of these captures, the plasma formation can be monitored and its relative density, temperature and stability can be inferred. Even though these are not absolute measurements and their accuracy is limited, this simple experimental set-up allows us to determine a range of optimum operation parameters for the ion source, in terms of the gas flow and RF power and frequency required to appropriately operate the source.

## 3. Experimental procedure and results

The direct CCD captures provide basic snapshots of the Hydrogen plasma luminescence, which can greatly vary depending on the RF power injected to the plasma chamber and the supplied Hydrogen mass flow. Once captured the images, it is pretty straightforward to obtain the relative intensity of the respective red, green and blue (RGB) color channels, as well as to calculate the global luminosity of each image. This luminosity can give an indirect measurement of the relative plasma density obtainable for each operational set of parameters in the ion source, which, in turn, will determine the maximum attainable beam current under these conditions.

In Figure 2 the the relative luminosities measured in the plasma chamber are shown, depending on the RF power and the Hydrogen gas mass flow delivered to the chamber. It is readily seen that higher gas flows correspond to higher plasma luminosities, provided enough RF power is provided. Also, for each flow regime, the luminosity increases almost linearly with the absorbed power once a minimum threshold value (around 50 W RF power) is crossed. These results are easily explainable and they provide an easy means of governing the plasma density in the chamber, and thus the proton beam current of the ion source, by appropriately regulating the RF power and Hydrogen flow.

If a low-cost diffraction grating (100 lines/mm) is mounted as a standard filter on the CCD camera lens, instead of capturing the direct image of the plasma and its luminosity, we can very easily perform a basic spectral analysis of the emitted light which can give us valuable information on the plasma formation.

[1] The amplifier has a nominal 100 W max. output when connected to a 50 Ohm load. Because there is no isolator between the amplifier and the plasma chamber the impedance seen by the amplifier is that of the plasma, which is not constant at 50Ohm, allowing the output power to be slightly larger than nominal.

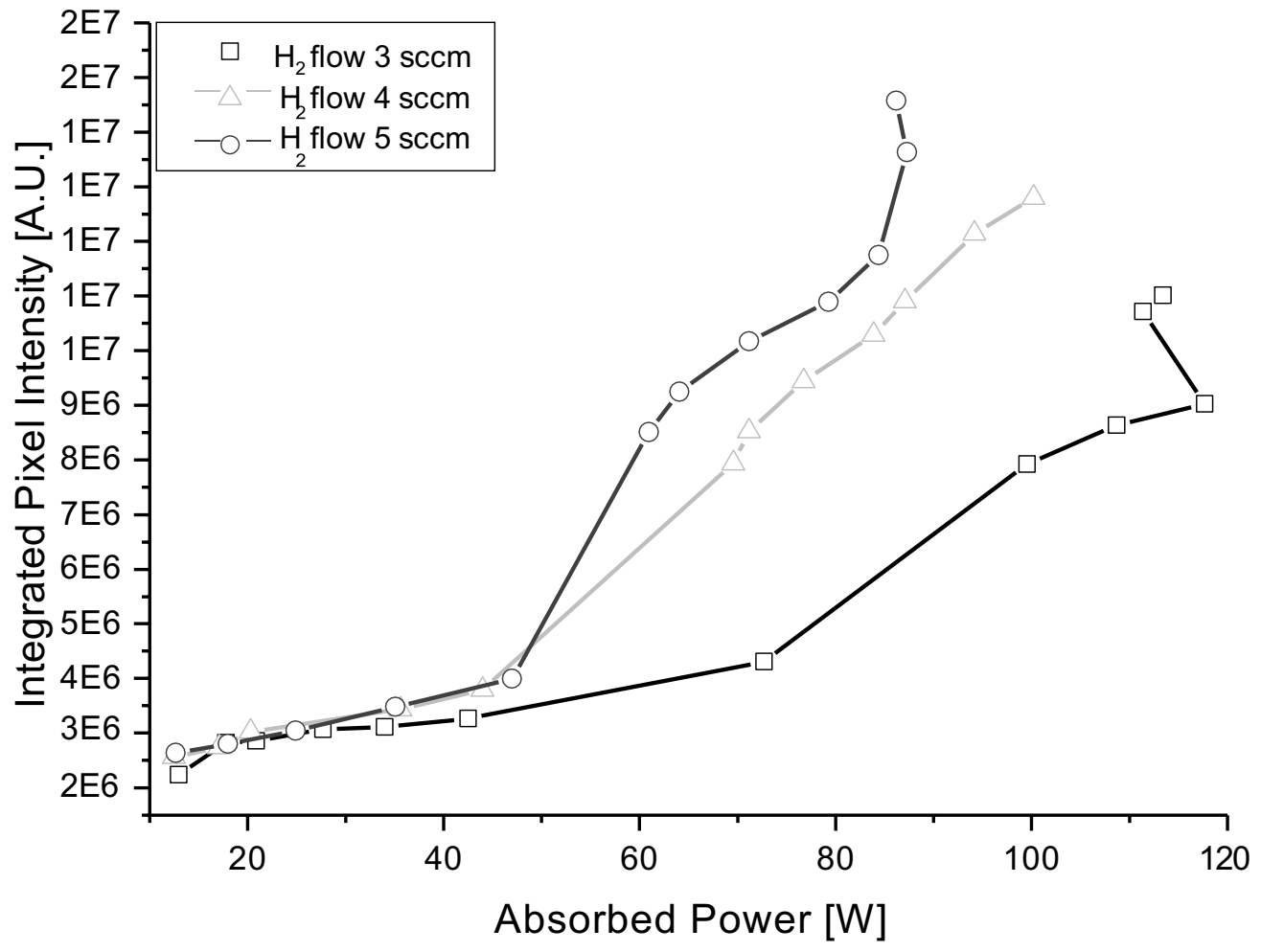

Figure 2: *Relative luminosity of the plasma CCD images as a function of the absorbed RF power in the plasma chamber, for different values of Hydrogen mass flow (3 GHz RF frequency)*

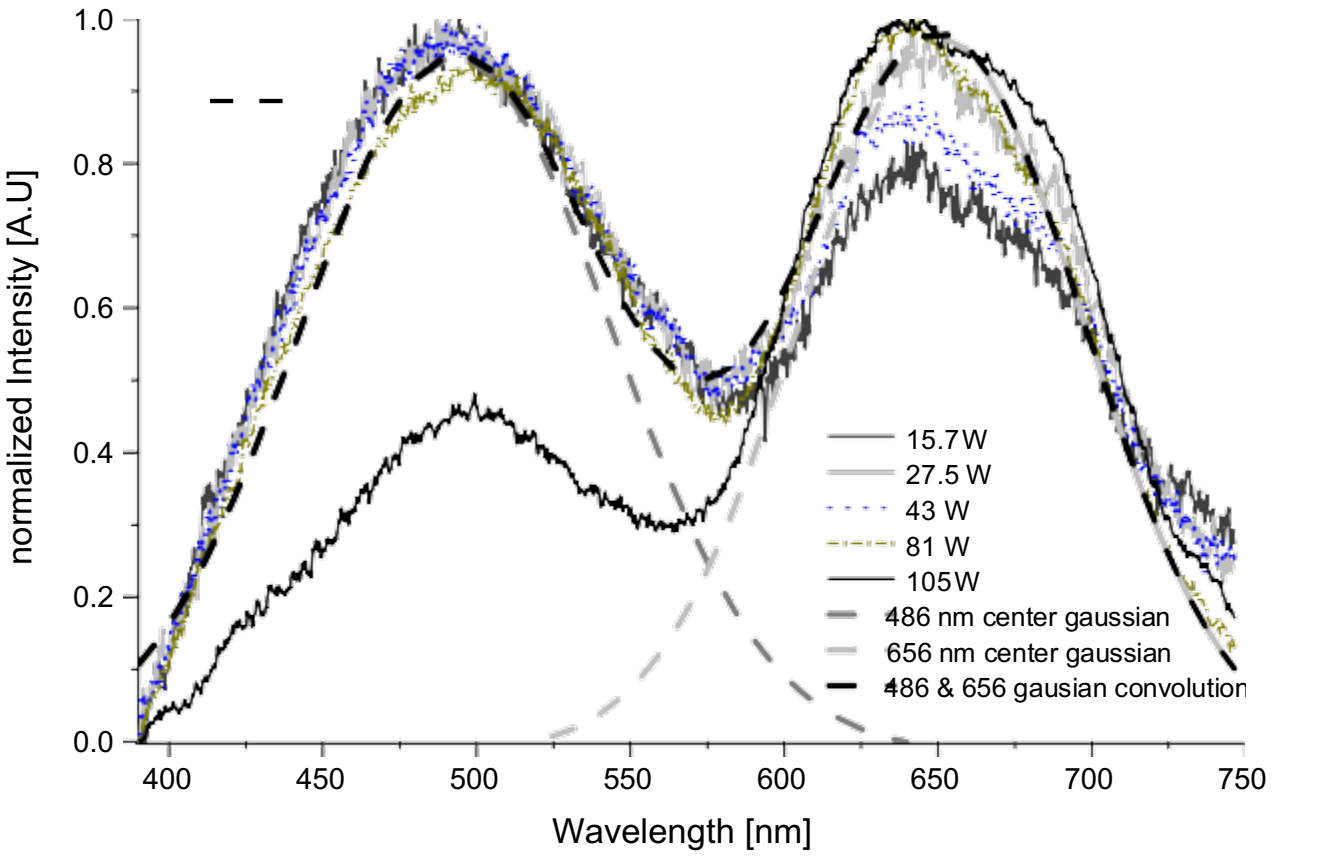

Figure 3: *Family of normalized plasma emission spectra directly measured with a diffraction grating for a range of 15.7 to 105 W of absorbed RF power and its gaussian fitting for the case corresponding to 93 W. (3 sccm Hydrogen mass flow, 3 GHz RF frequency)*

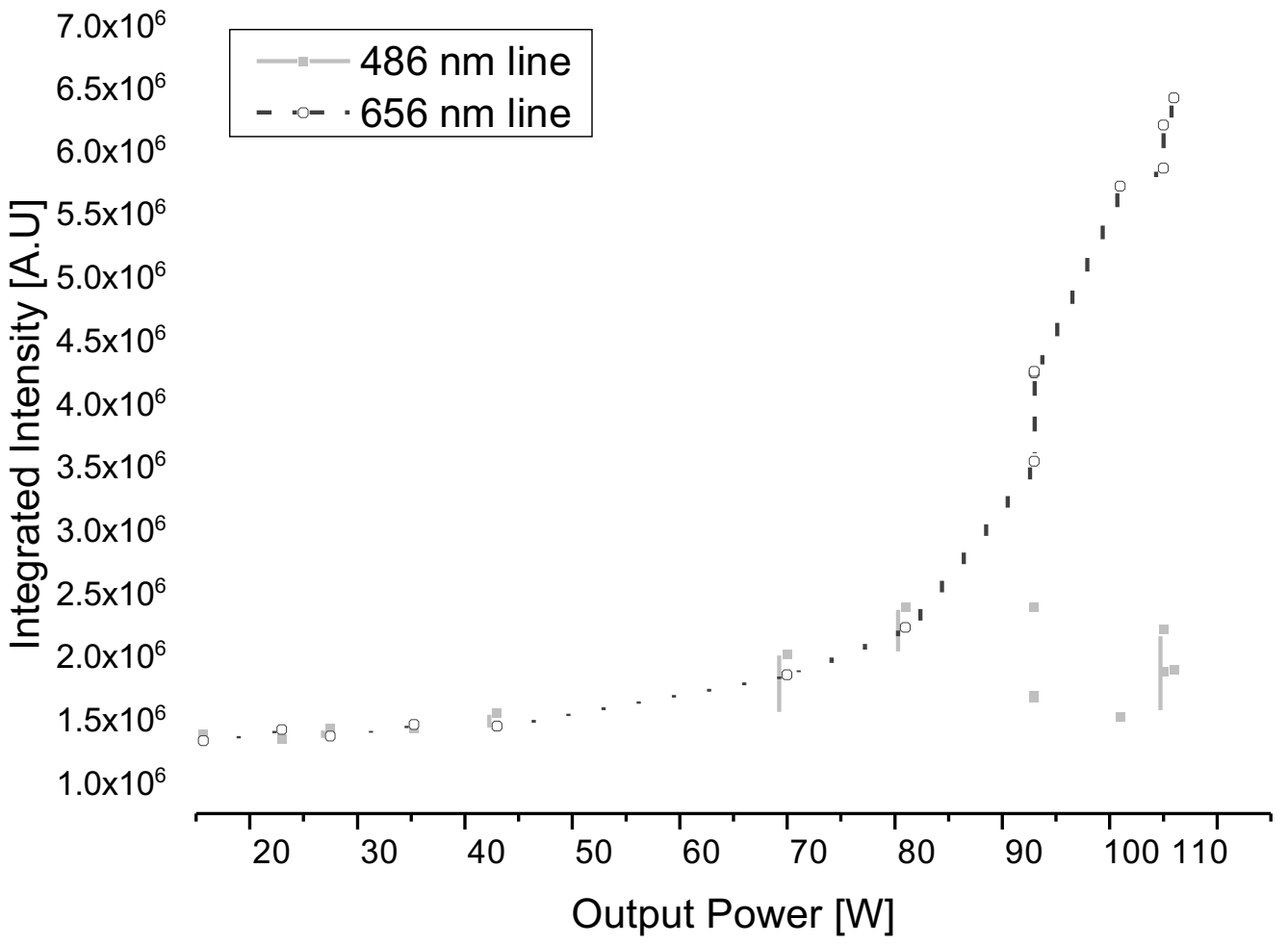

Figure 4: *Plasma emission intensities of $H_\alpha$ (656 nm) and $H_\beta$ (486 nm) as a function of the RF power absorbed in the plasma chamber. (3 sccm Hydrogen mass flow, 3 GHz RF frequency)*

Figure 3 shows a family of normalized emission spectra of the Hydrogen plasma formed in the ion source chamber as measured with the same CCD as above but now through the diffraction grating. The spectra were normalized because as the power was increased the absolute intensity of the peaks increased and the detector saturated, to compensate for this the exposure time had to be changed. Since we are only looking at the ratio of the two peaks present, having changed the exposure times doesn't present an issue. The measurement corresponds to 3 sccm Hydrogen mass flow and a range of absorbed RF power values from 15.7 to 105 W. An example of the gaussian fitting of the measured peaks is also shown as a dashed line in the same Figure for the spectrum corresponding to 93 W RF power. It is readily seen that in all cases the peaks are centered approximately around $\lambda$=656 nm $\lambda$=486 nm, corresponding, as expected, to the wavelengths of the Hydrogen spectral lines $H_\alpha$ and $H_\beta$ in the Balmer series.

The Hydrogen plasma formation can be characterized by representing the measured peak emission spectrum intensities as a function of the RF power delivered to the chamber for each gas flow. In Figure 4 the evolution of intensities of both emission peaks in terms of the gaussian fitting area corresponding to each one is represented. As shown, both of them show in general an increase in intensity as power increases, which correspond with the expected result of electron energy transition probability growth with RF power absorption.

At powers higher that 50 W, when electron energies increase, $H_\alpha$ shows a larger increase in intensity compared to $H_\beta$, so the inverse ratio between both intensities de-crease, as shown in Figure 5(due to the broad nature of the peaks, the area of the fitted gaussians is used to calculate the ratios). This behaviour is coherent with previous observations in Hydrogen and Argon plasmas found in the literature [7], and can be explained considering the fact that direct and dissociative excitation cross-section of $H_\alpha$ is higher compared to $H_\beta$, and in addition the cross-section increases as energy increments [8].

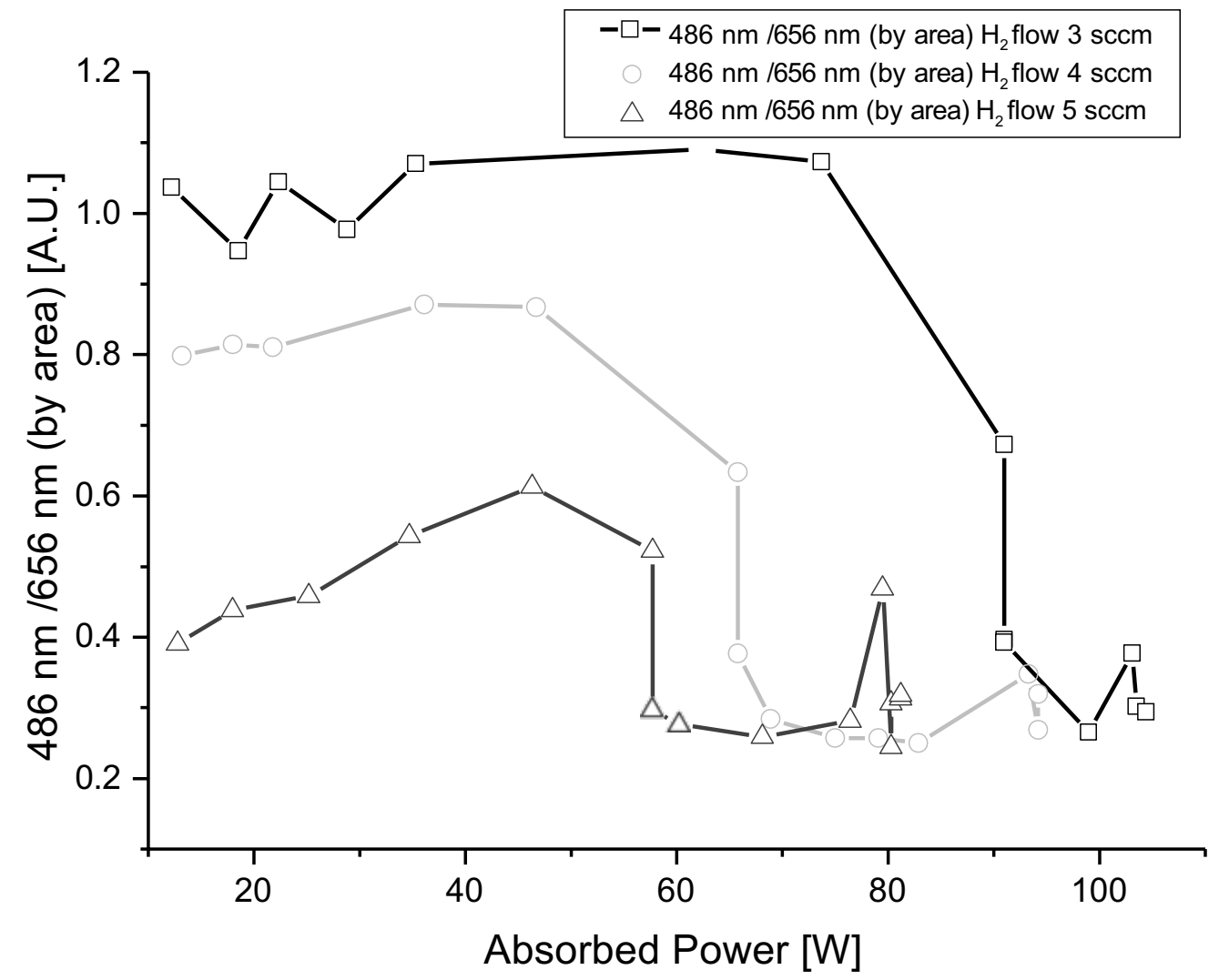

Figure 5: *Ratio of emission intensities $H_\beta$ / $H_\alpha$ as a function of the RF power absorbed in the plasma chamber for different values of Hydrogen mass flow (3 GHz RF frequency)*

As a general comment it is worth mentioning that these simple optical measurements allow us to basically characterize the plasma formation in the chamber as well as determining ranges of optimum parameters for the ion source operation. It looks clear from the above experimental results and their graphical representations, that higher plasma densities are obtained for higher Hydrogen flows and that for these higher flows (around 5 sccm) stable plasma production and steady density growth is achieved if RF powers above 60 W are used. This results give simple and easy to apply guidelines for the appropriate operation of the ion source.

## 4. Conclusions

A simple optical emission spectroscopy technique has been developed, and has proved to be useful to characterize the Hydrogen plasma obtainable in a new compact low-current ECRIS. The method, which is based on processing simple CCD captures of the plasma luminescence obtained both directly and through a low-cost 100 lines/mm transmission diffraction grating, has allowed us to characterize the ECRIS plasma production over the whole range of nominal RF power and available Hydrogen flow in the ion source, under a constant magnetic field, tuned to meet the ECR condition at 3 GHz.

It has been shown that if only limited accuracy is needed, optimal values for the ion source relevant operation parameters, including RF frequency, power and Hydrogen mass flow, can be easily obtained without the need of sophisticated equipment. This simple approach has allowed us to measure the relative spectral emission intensities $H_\alpha$ and $H_\beta$ in the Balmer series found in our ECRIS plasma, as well as to determine a region of optimum performance of the source, in terms of stable plasma formation and steady plasma density growth with the applied RF power.

For the low-current PIT30 ECRIS under study it has been observed that, as expected, the highest plasma densities are obtained for Hydrogen flows higher than 4 sccm, and is has also been seen that stable plasma production and steady density growth is achieved once a threshold of 60 W of RF power is overcome. These results provide an easy means of governing the plasma formation in the chamber, and thus the proton beam current of the ion source, by appropriately regulating the source operation parameters around the mentioned range.

As a whole, it has been shown that, despite its simplicity and low cost, the presented plasma characterization approach based on simple optical emission spectroscopy is easily implementable, it is very effective, and can well cope with basic plasma diagnostic tasks, at least as far as the successful operation of the ion source is concerned.

## Acknowledgements

The authors are grateful to the Department of Economic Development and Infrastructures of the Basque Government for partial support of this work under the Collaboration agreement with UPV/EHU in the field of Particle Accelerator Science and Technology.